\RequirePackage{lineno}
\documentclass[aps,showpacs,floatfix,twocolumn,byrevtex,superscriptaddress]{revtex4-1}

\usepackage{amsmath}
\usepackage{amssymb}
\usepackage{amstext}
\usepackage{amsopn}
\usepackage{amsfonts}
\usepackage{amsxtra}
\usepackage[english]{babel}
\usepackage{amsmath}
\usepackage{graphicx}
\usepackage{float}
\usepackage{bm}
\usepackage{multirow}
\usepackage{dcolumn}
\usepackage{hyperref}
\usepackage{CJK}
\usepackage[utf8]{inputenc}

\begin{document}

\title{Electronic correlations in the van der Waals ferromagnet Fe$_3$GeTe$_2$ revealed by its charge dynamics}
\author{M. Corasaniti$^{\dag}$}
\affiliation{Laboratorium f\"ur Festk\"orperphysik, ETH - Z\"urich, CH-8093 Z\"urich, Switzerland}
\author{R. Yang$^{\dag}$}
\affiliation{Laboratorium f\"ur Festk\"orperphysik, ETH - Z\"urich, CH-8093 Z\"urich, Switzerland}

\author{K. Sen}
\affiliation{Institut f\"ur Quantenmaterialien und -technologien, Karlsruher Institut f\"ur Technologie, D-76021 Karlsruhe, Germany}

\author{K. Willa}
\affiliation{Institut f\"ur Quantenmaterialien und -technologien, Karlsruher Institut f\"ur Technologie, D-76021 Karlsruhe, Germany}

\author{M. Merz}
\affiliation{Institut f\"ur Quantenmaterialien und -technologien, Karlsruher Institut f\"ur Technologie, D-76021 Karlsruhe, Germany}

\author{A. A. Haghighirad}
\affiliation{Institut f\"ur Quantenmaterialien und -technologien, Karlsruher Institut f\"ur Technologie, D-76021 Karlsruhe, Germany}

\author{M. Le Tacon}
\affiliation{Institut f\"ur Quantenmaterialien und -technologien, Karlsruher Institut f\"ur Technologie, D-76021 Karlsruhe, Germany}

\author{L. Degiorgi$^*$} 
\affiliation{Laboratorium f\"ur Festk\"orperphysik, ETH - Z\"urich, CH-8093 Z\"urich, Switzerland}

\date{\today}

\begin{abstract}
The layered van der Waals ferromagnetic Fe$_3$GeTe$_2$ harbours an unconventional interplay between topology and magnetism, leading to a large anomalous Hall conductivity at low temperatures. Here, we investigate the temperature dependence of its charge dynamics and reveal that upon entering the ferromagnetic state at $T_C \sim 200$ K and further lowering the temperature there is the onset of a gradual spectral weight reshuffling from the mid-infrared range towards far- as well as near-infrared frequencies. This two-fold spectral weight transfer indicates the important role of the Hund's coupling as primary source for electronic correlations and signals an incoherent-coherent crossover at low temperatures. Our findings also convey the electronic environment, based on nodal-line topological states, favouring the large anomalous Hall conductivity.
\end{abstract}

\maketitle

The interplay of lattice symmetry and topology has been ever since considered of paramount importance for the overall understanding of the peculiar electronic properties in topological materials \cite{Hasan2010,Qi2011,Ando2015,Chiu2016,Bansil2016,Armitage2018}. Recently, the scientific interest, both experimental and theoretical, moved towards exploring the mutual impact of topology and magnetism, which remains in many aspects poorly understood. For instance in the case of ferromagnetic (FM) topological Weyl systems with broken time reversal symmetry, an enhanced, large anomalous Hall effect (AHE), i.e., an antisymmetric contribution to the off-diagonal resistivity linked to the magnetisation rather than the applied magnetic field, may be foreseen and is of purely intrinsic nature, fully related to the geometrical properties of the electronic structure and not to the extrinsic impurity scattering effects \cite{Burkov2014}. 

In this context, several layered van der Waals (vdW) materials \cite{Zeng2018} have been identified as topological key-players \cite{Wang2019}.  They also encounter Fe$_3$GeTe$_2$, one of the rare and elusive FM topological semimetals with a high Curie temperature $T_C \sim$ 200 K \cite{Chen2013,May2016}. The vdW materials are considered as front-runners of two-dimensional (2D) electronic systems with weak interlayer coupling, which facilitates exfoliation into few-layered nanosheets. The 2D quantum confinement of their electronic properties makes them of interest for potential applications in the area of electronic and spintronics devices \cite{Deng2018,XWang2019}.

We wish to highlight three major findings, drawn from the extensive research activity on the title vdW compound. First of all, the electronic structure, established by first-principles calculations, corroborates the presence of an orbital-driven nodal-line in Fe$_3$GeTe$_2$, connecting two orbitals in adjacent layers \cite{Kim2018}. Concomitantly, this is the prerequisite, via the impact of spin-orbit coupling (SOC), in order to induce so-called large Berry phase curvature, which may lead to a considerable AHE \cite{Kim2018,Wang2017}. Secondly, electronic correlation effects were conjectured in this ferromagnet, together with a substantially large effective electron mass renormalisation \cite{Zhu2016}. Third, the electronic band structure imaged by angle-resolved photoemission spectroscopy (ARPES) reveals spectral weight redistribution at the FM transition, ascribed to the exchange splitting \cite{Zhu2016,Zhang2018}. The study of Ref. \onlinecite{Zhang2018} moreover advances the gradual emergence of a Kondo lattice behaviour with the presence of heavy electrons upon lowering the temperature ($T$) below $T^*\sim$ 100 K. Indeed, the specific heat data and the $dc$ transport properties also convey a large Sommerfeld value and the presence of an undoubtedly change of slope at $T^*$, respectively, which are typical fingerprints of heavy-fermion systems \cite{Zhu2016,Zhang2018,Degiorgi1999,Basov2011}. All this led to speculate the interplay between ferromagnetism and heavy electron states in a 3$d$ material \cite{Zhang2018,Zhao2020}. However, the scenario, based on the itinerant Stoner model \cite{Zhu2016,Zhang2018}, has been opened to debate by the latest ARPES data \cite{Xu2020}, which point out a massive enhancement of the quasiparticles coherence below $T_C$, following a suppression of magnetic fluctuations, and advance the crucial role played by local magnetic moments. 

A challenging goal ahead is to possibly unify such a vast wealth of knowledge and physical concepts within a comprehensive experimental set of data, covering a broad spectral range as well as addressing all relevant $T$ intervals across $T_C$ and $T^*$. Here, we present a thorough optical investigation as a function of $T$ which addresses the complete charge dynamics. We can image the $T$ evolution of the electronic properties and extract the nature of their correlations; first by uniquely disentangling various relevant parameters and quantities, like the scattering rate and the plasma frequency within the intraband contribution of the absorption spectrum, and then by accessing the interband excitations spanning the energy interval from the Fermi level ($E_F$) to states deep into the electronic structure. We discover a two-fold spectral weight transfer from the mid-infrared towards lower as well as higher energy scales and developing within an energy interval of about 2 eV. This bears testimony to electronic correlations driven by Hund's coupling and to the presence of an incoherent-coherent crossover at low $T$, likely underlining the enhancement of the effective electronic dimensionality. Furthermore, the excitation spectrum reveals the absorption feature associated with the SOC-gapped nodal-line and thus the ingredients of the electronic environment necessary for the large AHE.

The reflectivity $R(\omega)$ of a Fe$_{3}$GeTe$_{2}$ single crystal was collected as a function of $T$ from the far infrared (FIR) to the ultraviolet at nearly normal incidence with respect to the \textit{ab}-plane of the stacked Fe$_3$GeTe$_2$ bilayer \cite{Dressel2002}. This is the prerequisite in order to perform a reliable Kramers-Kronig transformation of $R(\omega)$, giving access to all optical functions. We refer to the Supplemental Material, Ref.~\onlinecite{SM} for further details. 

\begin{figure}[tb!]
\centerline{
\includegraphics[width=1\columnwidth]{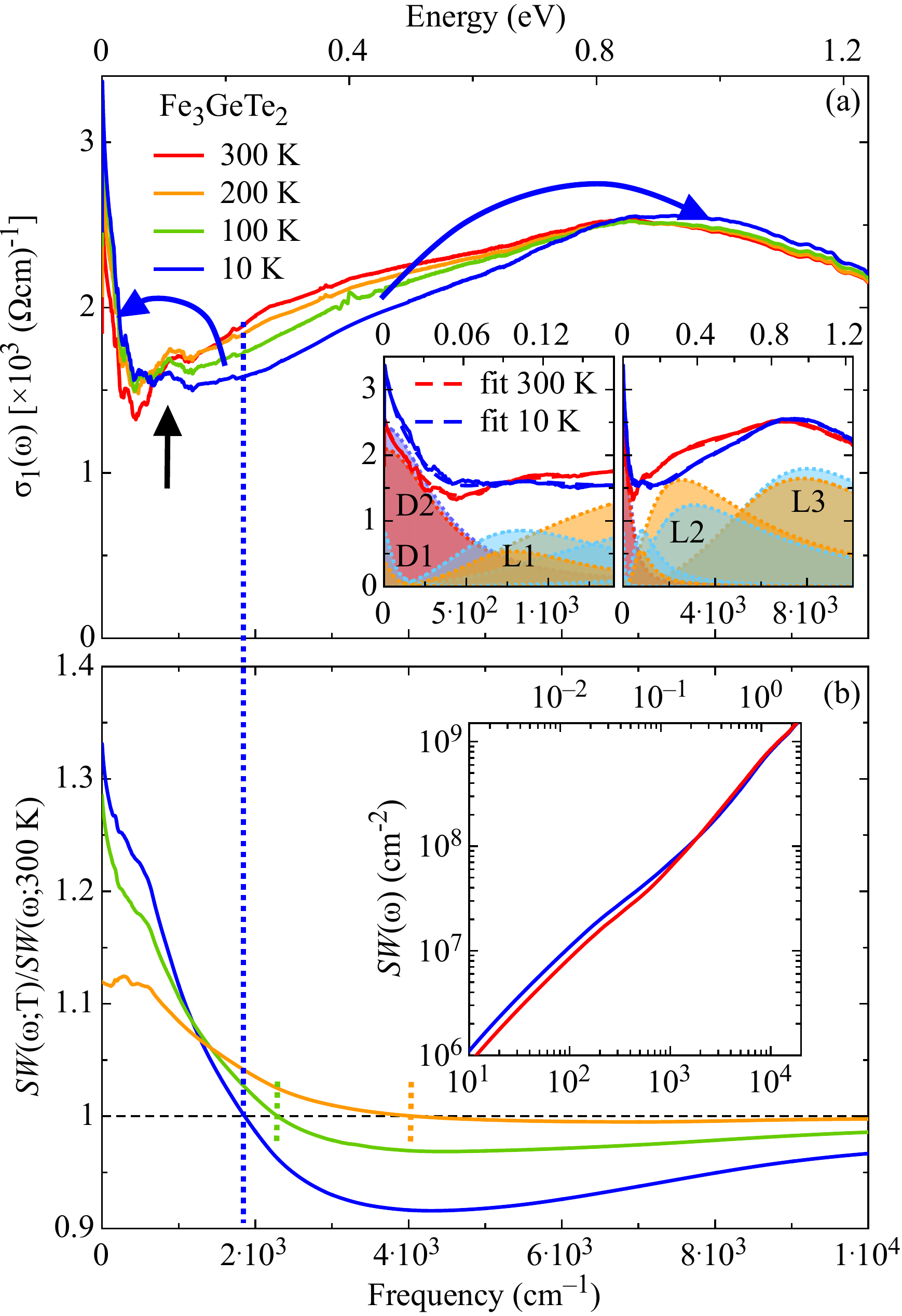}
}
\caption{(a) Real part ($\sigma_1(\omega)$) of the optical conductivity of Fe$_3$GeTe$_2$ as a function of $T$ in the spectral range below 10$^4$ cm$^{-1}$ (1 eV = 8.06548$\times$10$^3$ cm$^{-1}$). The insets are its blow-up in the FIR-MIR and MIR-NIR spectral range at 300 and 10 K, together with the Drude-Lorentz fit (Eq. S1 in Ref. \onlinecite{SM}) and its constituent components: Di (i=1 and 2) are the narrow and broad Drude terms, while Li (i=1 to 3) are the Lorentz harmonic oscillators (Fig. S4 in Ref. \onlinecite{SM}). Reddish and bluish colors refer to 300 and 10 K, respectively. The black arrow in the main panel points out the absorption at $\sim$ 800 cm$^{-1}$ (see text). (b) Integrated spectral weight ($SW$) as a function of frequency at 10, 100 and 200 K (Eq. \ref{eq_spectralweight}), normalised by its value at 300 K. The inset shows the bare $SW$ quantity at 10 and 300 K. The dotted, vertical blue line across both panels divides the investigated spectral range into two intervals, for which the reshuffling of $SW$ at 10 K occurs towards low (left hand-side) and high (right hand-side) energy scales, respectively (as schematically indicated by the (rounded) blue arrows in panel (a)). At this energy scale $\omega^*$ (see text), $SW(\omega^*; T)/SW(\omega^*; 300 K)$ = 1. The dotted, vertical orange and green segments in panel (b) define the corresponding $\omega^*$ at 200 and 100 K, respectively.} 
\label{sigma1_SW}
\end{figure}

We start off with the real part ($\sigma_1(\omega)$) of the optical conductivity, shown in Fig. \ref{sigma1_SW}(a) at four selected $T$ in the energy interval spanning the FIR, mid (MIR) and near (NIR) infrared spectral ranges. There is a narrow metallic component which gets stronger upon lowering $T$ (i.e., it gains spectral weight, see below) and which merges into a broad MIR feature peaked at about 7000 cm$^{-1}$. The low frequency side of this MIR absorption is depleted, while its high energy one is enhanced upon lowering $T$, respectively. Before going any further, we draw the attention on the additional absorption located around 800 cm$^{-1}$ (black arrow in Fig. \ref{sigma1_SW}(a)), to which we will get back towards the end of this paper. 

Below 10$^4$ cm$^{-1}$, we thus uncover a substantial $T$ dependence of the optical response, which underlines a reshuffling of spectral weight ($SW$). In order to elucidate the $SW$ redistribution, we introduce the so-called integrated $SW$ of the measured $\sigma_1(\omega)$ up to well-defined cut-off energies ($\omega$), which is given by:

\begin{equation}
\label{eq_spectralweight}
SW(\omega;T) = \frac{Z_0}{\pi^2}\int_0^{\omega}\sigma_1(\omega'; T)d\omega',
\end{equation}
expressed in units of cm$^{-2}$ ($Z_0$ = 376.73 $\Omega$, being the impedance of free space) \cite{Dressel2002}. This model-independent quantity is related to the number of the effective carriers (normalized by their effective mass) contributing to the optical processes up to $\omega$ and images the evolution of the electronic band structure upon varying $T$. Therefore, in the $\omega\rightarrow\infty$ limit, it is expected to merge to a constant value at all \textit{T}, satisfying the optical $f$-sum rule \cite{Dressel2002}. The full recovery of $SW$ in our data is already achieved at energies of about 1.5 eV from $E_F$ (inset of Fig. \ref{sigma1_SW}(b)). For the purpose of our discussion, we consider the ratio $SW(\omega;T)/SW(\omega; 300 K)$ which emphasises the relevant energy scale of the $SW$ transfer as a function of $T$ with respect to 300 K. If there is a transfer of $SW$ from high to low energies, the $SW$ ratio will exceed 1 at low energies and then smoothly approach 1 upon increasing $\omega$ until the full energy scale of the low-energy resonance is reached. For instance, $SW$ may move into the metallic (Drude) zero-energy mode. If there is a transfer of $SW$ from low to high energies though, the $SW$ ratio will fall below 1 until the total energy scale of $SW$ transfer is reached. This latter case may suggest some erosion of density-of-states (DOS), as it would occur by an electronic bands reconstruction. Figure \ref{sigma1_SW}(b) displays the $SW$ ratio at $T < $ 300 K. It clearly emphasises the two-fold $SW$ reshuffling to low as well as to high energies, which gets more pronounced upon decreasing $T$. The dotted, vertical line across both panels in Fig. \ref{sigma1_SW} marks the energy scale (e.g., $\omega^*\sim$ 2000 cm$^{-1}$ at 10 K) differentiating those two distinct directions of the $SW$ redistribution (as mimicked by the rounded blue arrows in Fig. \ref{sigma1_SW}(a)). From Fig. \ref{sigma1_SW}(b), one can readily observe that $\omega^*$ moves to higher values upon increasing $T$. The low energy $SW$ accumulation thus occurs in a narrower energy interval from $E_F$, while the $SW$ removal and reshuffling towards high energies happens over a broader energy interval upon lowering $T$, respectively. 

\begin{figure}[tb!]
\centerline{
\includegraphics[width=1\columnwidth]{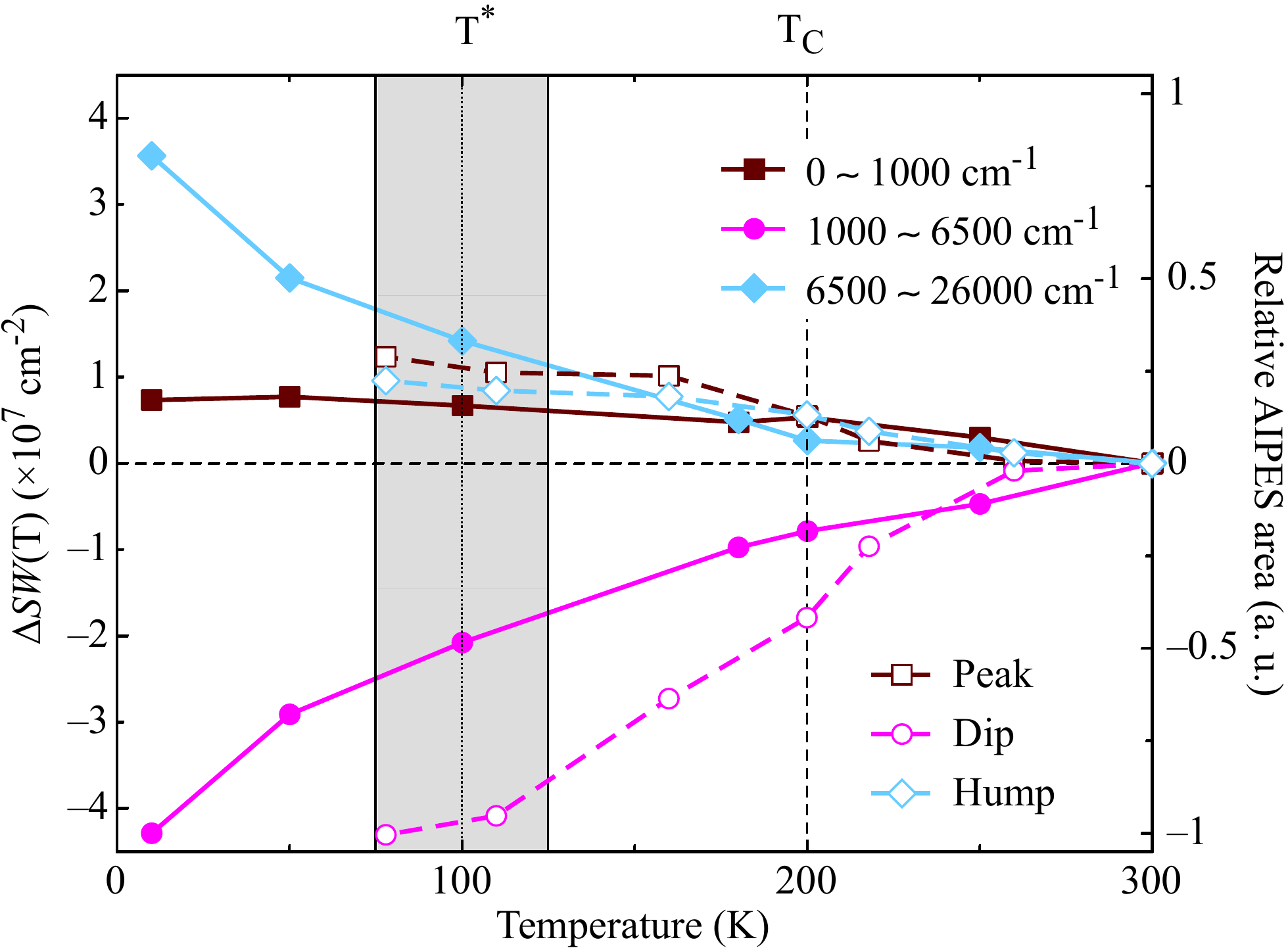}}
\caption{Relative spectral weight variation $\Delta SW(T) = SW(T) - SW(300 K)$ at 10, 100 and 200 K for three selected spectral ranges ($\tilde{\omega_1}$, $\tilde{\omega_2}$). Additionally, the relative areas of the peak, dip and hump (see text), observed in the $T$ dependence of the angle-integrated photoemission spectroscopy (AIPES), are reproduced from Ref. \onlinecite{Zhang2018}. The relative areas were renormalised by the dip area at 80 K.} 
\label{SW-distribution}
\end{figure}

In order to further elaborate on the $T$ dependence of the $SW$ distribution, it is instructive to identify relevant $T$ ranges with respect to the $dc$ transport properties (Fig. S2 in Ref. \onlinecite{SM}). There are two main $T$ regimes: around $T_C$ at the FM transition and the crossover interval at $T^*$ (dashed line and grey shaded area in Fig. \ref{SW-distribution}, respectively). We then generalise Eq. \ref{eq_spectralweight} as integral of $\sigma_1(\omega)$ between $\tilde{\omega_1}$ and $\tilde{\omega_2}$, and calculate the relative $SW$ variation $\Delta SW(T) = SW(T) - SW(300 K)$ for appropriately selected energy ranges ($\tilde{\omega_1}$, $\tilde{\omega_2}$). Guided by the $T$ evolution of $\sigma_1(\omega)$ (Fig. \ref{sigma1_SW}(a) and Fig. S3(b) in Ref. \onlinecite{SM}), we consider the ($\tilde{\omega_1}$, $\tilde{\omega_2}$) ranges (0, 1000), (1000, 6500) and (6500, 26000) cm$^{-1}$, which we denote for brevity as FIR, MIR and NIR intervals, respectively. $\Delta SW(T)$ is shown in Fig. \ref{SW-distribution}; while the overall $SW$ in the MIR interval continuously decreases, it progressively increases in the FIR and NIR ones with decreasing $T$. Interestingly enough, the trends of $\Delta SW(T)$ at MIR and NIR intervals gradually evolve below $T_C$ and are even more consolidated when crossing $T^*$. The increase of $\Delta SW(T)$ at FIR energies upon lowering $T$ expires at $T < T^*$, where it saturates. Complementary, we analyse our data within the phenomenological Drude-Lorentz approach \cite{SM}.
This allows us focusing the attention on the $SW$ reshuffling within selected spectral ranges (i.e., related to each fit component, Fig. S4 in Ref. \onlinecite{SM}). The insets of Fig. \ref{sigma1_SW}(a) confirm indeed that upon lowering $T$ the overall metallic (Drude) contribution and its high frequency tail represented by the FIR harmonic oscillator (HO) L1 as well as the high frequency NIR HO L3 gain $SW$ (corresponding to the color shaded areas underneath each fit component \cite{SWDL}), while $SW$ of MIR HO L2 is definitely depleted (see also Fig. S7 in Ref. \onlinecite{SM}).

The $SW$ redistribution evidenced from the optical response images a remarkable reconstruction of the electronic bands structure over a broad energy interval upon lowering $T$. There is indeed an immediate link with the recent ARPES findings \cite{Zhang2018,Kim2018,Xu2020}. Particularly in the angle-integrated photoemission spectroscopy (AIPES) spectra of Ref. \onlinecite{Zhang2018} several features evolve upon lowering $T$ below $T_C$: a peak centred at 50 meV, a dip around 200 meV and a hump at 500 meV from $E_F$. The peak and hump convey a sizeable DOS related to the (relatively flat) electronic bands $\gamma$ close to the K point of the Brillouin zone (BZ) and $\eta$ as well as $\epsilon$ at K and $\Gamma$ point of BZ, respectively \cite{Zhang2018}. The absorptions in $\sigma_1(\omega)$ at $\sim$ 0.1 and 0.5-0.7 eV are due to interband transitions from those $\gamma$ as well as $\epsilon$ and $\eta$ bands into empty states above $E_F$. The $T$ dependence of the relative area of those AIPES features was attributed to a spectral weight transfer \cite{Zhang2018}, which in rather common in FM materials \cite{Kim1992,Okabayashi2004,Takahashi2001} and agrees astonishingly well with the optical results, as shown in Fig. \ref{SW-distribution}. 

The two-fold transfer of the optical $SW$ from the MIR to NIR as well as FIR energies hints to the intricate electronic properties in the presence of both itinerant and localised electrons from different Fe 3$d$ orbitals. Such an orbital-selective behavior, which is widely observed in multi-orbital systems like iron-based superconductors \cite{deMedici2014,Yi2013,Yi2015}, is reinforced by the FM order in Fe$_3$GeTe$_2$ and calls for mechanisms going beyond the purely itinerant (Stoner) approach or localised (Heisenberg) scenario. A distribution of $SW$ into low (coherent) and high (incoherent) energy components of the optical response, strengthened upon lowering $T$, may be expected within a Mott-Hubbard approach for strongly correlated metals \cite{Rozenberg1995,Si2009}. In such a case, the related $SW$ redistribution towards high energies occurs at energy scales of the order of the Hubbard $U$, as observed experimentally \cite{Rozenberg1995,Qazilbash2008,Qazilbash2009}. $U$ in Fe$_3$GeTe$_2$ is of about 5 eV \cite{Zhu2016,Kim2018,Zhang2018}, which implies much higher energy scales for the $SW$ reshuffling than observed (Fig. \ref{sigma1_SW}), thus making this scenario unlikely. On the other hand, the $SW$ shift to energies less than 1.5 eV (inset of Fig. \ref{sigma1_SW}(b)) favours a mechanism related to the Hund's coupling \cite{Georges2012}, as envisaged \cite{Johannes2009,Yin2011} for as well as observed \cite{Wang2012,Schafgans2012} in ferropnictides. Upon switching off thermal excitations with lowering $T$, a fraction of less itinerant Fe 3$d$ electrons will be progressively localised via Hund's coupling by the local 3$d$ moments of the FM state, leading to $SW$ transfer at energies higher than the Hund's coupling itself. In this context, it is worth mentioning that an exchange (Hund's) coupling $J \sim$ 0.79 - 0.9 eV, so compatible with the energy scale of the $SW$ transfer (Figs. \ref{sigma1_SW} and \ref{SW-distribution}), was found to be appropriate for the title compound \cite{Kim2018,Zhu2016}. 

We now turn our attention to the $SW$ shift towards low energy scales upon decreasing $T$. Below $T_C$ with suppressed magnetic fluctuations, part of the removed MIR $SW$ accumulates into the effective metallic contribution of $\sigma_1(\omega)$ (insets of Fig. \ref{sigma1_SW}(a) and Fig. \ref{SW-distribution}) and signals a so-called incoherent-coherent crossover with decreasing $T$, which has been observed on several occasions in superconducting and, quite generally, in strongly correlated materials like heavy-fermion (Kondo) systems \cite{Wang2002,Yang2017,Jonsson2007,Degiorgi1999,Basov2011}. Because of the multi-band nature of Fe$_3$GeTe$_2$, we consider two Drude terms in our phenomenological fit of the effective metallic component in $\sigma_1(\omega)$ (insets of Fig. \ref{sigma1_SW}(a) and Fig. S4 in Ref. \onlinecite{SM}). Interestingly enough, the narrow one undergoes proportionally the largest increase, corresponding to an enhancement by a factor of 2.4 of its $SW$ from 300 down to 10 K (in contrast to a factor of about 1.1 for the broad Drude term, Fig. S5(a) in Ref. \onlinecite{SM}). We propose to consider the narrow Drude term as representative of the most coherent component in $\sigma_1(\omega)$, since it strongly develops at low $T$. Therefore, Hund's coupling plays again a central role so that it causes not only moment formation at high energy but, through the incoherent-coherent crossover, it may also be responsible for the low energy correlation effects in the low $T$ metallic phase \cite{Haule2009,Yin2012}, imaged by a large effective electron mass renormalisation \cite{Zhu2016,SM}. The enhanced coherence upon lowering $T$ is also reflected in the (Drude) scattering rates (Fig. S5(b) in Ref. \onlinecite{SM}). 

The low energy $SW$ shift, effectively underlining the enhanced coherence at low $T$, may be considered as smoking gun evidence for a progressive dimensionality crossover in the layered Fe$_3$GeTe$_2$ semimetal. While optical investigations along the $c$-axis (Fig. S1(b) in Ref. \onlinecite{SM}) would substantiate this speculation, the presence of coherent charge carriers at low $T$ conjugates with the development of a stronger interlayer coupling (as shown in a variety of low-dimensional metals \cite{Vescoli1998,Valla2002}), also facilitated here by the suppression of magnetic fluctuations and thus of scattering channels at $T < T_C$ \cite{SM}. This would pair with the observed $T$ dependence of the $dc$ transport properties along the less conducting $c$-axis \cite{Kim2018,Wang2017}, which indeed changes its slope in a more obvious fashion at $T^*$ than within the $ab$-plane \cite{dc}. This latter aspect turns out to be a generic feature of layered strongly correlated metals \cite{Valla2002}. 

\begin{figure}[tb!]
\centerline{
\includegraphics[width=1\columnwidth]{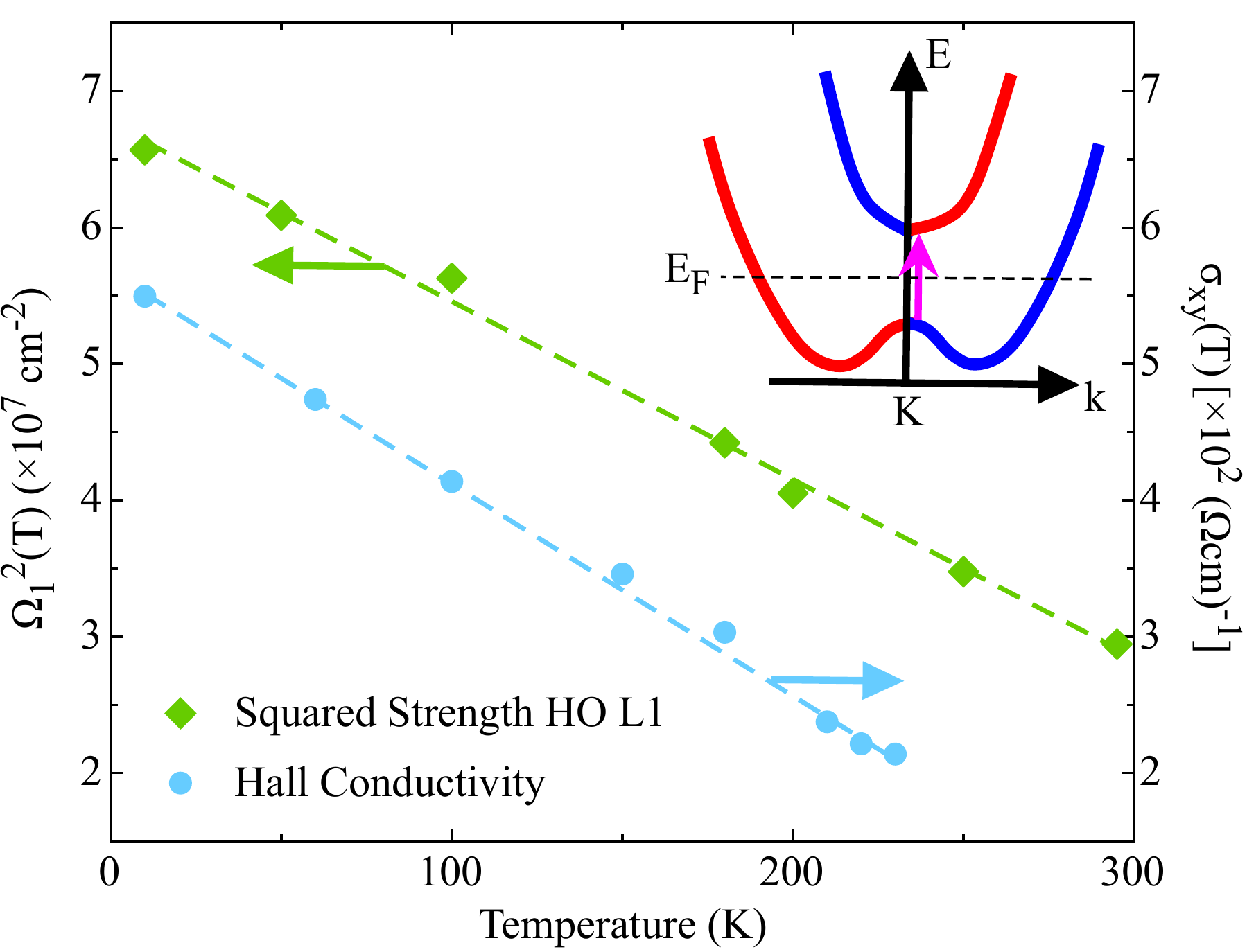}}
\caption{Temperature dependence of the squared strength ($\Omega_1^2$) of HO L1 \cite{SM,SWDL} and of the Hall conductivity $\sigma_{xy}$, reproduced from Ref. \onlinecite{Kim2018}. $\sigma_{xy}$ is read at the magnetic field $H$ = 9 T ($H \parallel c$-axis), where it saturates. It is known that $\sigma_{xy} \sim \sigma_{xy}^A$ in Fe$_3$GeTe$_2$ \cite{Kim2018,Wang2017}. The linear fits (dashed lines) are guide to the eyes. Inset: schematic sketch of the SOC-gapped band at the K-point of BZ. Red and blue refer to different combinations of the orbital degrees of freedom within the Fe$_3$GeTe$_2$ bilayer structure \cite{Kim2018}. The pink arrow indicates the excitation represented by HO L1 in our Drude-Lorentz fit \cite{SM}. } 
\label{AHE}
\end{figure}

In addition to the $SW$ transfer, discussed so far within the Hund's coupling framework, also HO L1 at $\sim$ 800 cm$^{-1}$ (Fig. \ref{sigma1_SW}(a), and Figs. S4 and S6 in Ref. \onlinecite{SM}) experiences a $SW$ enhancement by a factor $\sim$ 2 between 300 and 10 K (Fig. \ref{AHE} as well as Fig. S7 in Ref. \onlinecite{SM}). As said above, this absorption relates to the electron-like $\gamma$ band, observed in the intensity plot of the ARPES data at the K point of BZ \cite{Zhang2018} and resulting from the interlayer hybridisation for the stacked Fe$_3$Ge slabs \cite{Kim2018}. This latter two-fold degenerate state extends along the KH line in BZ and the degeneracy of such a nodal-line can be lifted by SOC with $E_F$ lying somehow in-between (inset of Fig. \ref{AHE}), as shown by first-principles calculations exclusively for a FM order along the $c$-axis \cite{Kim2018}. Therefore, by inspecting the band structure near $E_F$ we believe that HO L1 mimics the dipole-active excitation across the SOC-gapped nodal-line, also called avoided-band crossing. Such a topological state triggers a large Berry curvature and is supposed to act as a 1D vortex line that generates Berry flux in the momentum space, the latter being the prerequisite for AHE \cite{Kim2018}. In fact, the anomalous Hall conductivity ($\sigma_{xy}^A$) is proportional to the BZ integral of the convolution of the Berry curvature with the $T$ dependent occupation function \cite{Nagaosa2010,Weng2015}. Figure \ref{AHE} finally conveys our claim that the observed increasing strength ($\Omega_1$) in HO L1 upon lowering $T$ indeed indicates the increasing occupation of the valence band at the K point of BZ \cite{JDOS}, which further contributes to promote the enhancement of $\sigma_{xy}^A$ \cite{Kim2018,Wang2017}. We refer to Fig. S8 in Ref. \onlinecite{SM} for a complementary comparison between $\Omega_1$ and $\sigma_{xy}$.

In conclusion, the $T$ dependence of the charge dynamics in the vdW FM Fe$_3$GeTe$_2$ semimetal provides signatures of the dual itinerant and localised character of the Fe 3$d$ electrons and reveals a two-fold reshuffling of $SW$, signalling the role played by Hund's coupling as well as coherence effects. Furthermore, the gradual increase of $SW$ associated with the excitation involving the topological nodal-line gapped by SOC may account for the remarkable AHE upon lowering $T$. This testifies the unique interplay of FM and topology, which here depends ultimately on the environment of the Fe atom, tuneable by parameters affecting the design of the Fe$_3$Ge slabs. In a broad and generic context, this may shed new light on related 2D materials and foster ways for future manipulations of their AHE and its 2D quantum limit, in view of novel electronic functionalities. 

\section*{Acknowledgements}
The authors thank C.C. Le for fruitful discussions. This work was partially supported by the Swiss National Science Foundation (SNSF). K.S. acknowledges 'Networking Grant' from the Karlsruhe House of Young Scientists. K.W. acknowledges funding from the Alexander von Humboldt foundation. The contribution from M.M. was supported by the Karlsruhe Nano Micro Facility (KNMF).\\

$^{\dag}$ Both authors contributed equally to this work.\\

$^*$ Corresponding author: degiorgi@solid.phys.ethz.ch\\


\begin{thebibliography}{57}
\bibitem [{Vmm()}]{Vmmolar}%
  \BibitemOpen
  \href@noop {} {}\bibinfo {note} {The molar volume of Fe$_3$GeTe$_2$ is equal
  to 67.75x10$^{-6}$ m$^3$/mole \cite{Vm}.}\BibitemShut {Stop}%
\end{thebibliography}
%

\end{document}